\documentstyle[11pt,aaspp4]{article} 
\begin{document}
\title{Pulsar Emission and Force-Free Electrodynamics} 
\author{Andrei Gruzinov}
\affil{CCPP, Physics Department, New York University, 4 Washington Place, New York, NY 10003}

\begin{abstract}

Pulsar emission should primarily come from the magnetic separatrix... Combining theory and observations, we show that force-free electrodynamics (FFE) gives an accurate description of the large-scale electromagnetic field in the magnetospheres   of Crab-like pulsars. A robust prediction of FFE is the existence and stability of a singular current layer on the magnetic separatrix. We argue that most of the observed pulsar emission comes from this singular current layer.

\end{abstract}

\section{Introduction}

Two different things are done in this paper. First, we show that simple and pure theory -- force-free electrodynamics (FFE) -- provides an accurate quantitative description of the large-scale electromagnetic field in pulsar magnetospheres. Second, we use certain FFE results to propose that pulsar emission is mostly generated in the vicinity of the magnetic separatrix, which is the surface separating the regions of closed and open magnetic field lines. 

FFE describes electromagnetic fields in plasmas, but the only degrees of freedom in FFE are the electromagnetic fields of a special force-free geometry. The plasma is described implicitly. FFE makes no predictions regarding plasma emission. 

Stationary FFE equations describing pulsar magnetospheres have recently been solved (Contopoulos, Kazanas, Fendt 1999, Gruzinov 2005, 2006, Spitkovsky 2006). The shape of the pulsar magnetosphere, the singularity structure, and the spin-down power have been calculated (in a sense, see \S3). But these results hang in the air; being pure FFE results, they make no directly testable predictions \footnote{One may even say that FFE does make one prediction -- the spin-down rate should scale as the cube of the spin. This is known to be wrong. But no satisfactory explanation of the anomalous breaking of pulsars exists today. We will not count this as an argument against FFE.}. 

We think that FFE has actually found the primary site of pulsar emission. We propose that pulsar emission is due to a small dissipation effect on top of the ideal dissipationless FFE. The dissipation is basically Joule heating. The current layer on the separatrix consists of counter-propagating beams of electrons and positrons of formally infinite densities. For any ``reasonable'' anomalous resistivity, the Joule heating in the singular current layer exceeds the Joule heating in the bulk. Then the singular current layer should also dominate the pulsar emission. Our theory is obviously incomplete -- we do not calculate the emitted power spectrum and the light curves. But the case for the magnetic separatrix being the primary source of pulsar emission looks strong.

\section{FFE}

Here we formulate FFE and discuss its meaning and conditions of applicability. Then we show that FFE provides a valid description of the large-scale electromagnetic field in the magnetospheres of Crab-like pulsars.

In FFE, as in all plasma models, Maxwell equations are used for the electromagnetic field: 
\begin{equation}\label{maxwell}
\partial _t{\bf B} =-\nabla \times {\bf E},~~~ \partial _t{\bf E}=\nabla \times {\bf B}-{\bf j}.
\end{equation}
To close this system, one needs to calculate the plasma current density ${\bf j}$. In standard plasma physics this is done by solving the equations of motion of charged particles. In FFE one postulates the following Ohm's law:
\begin{equation}\label{ohm}
{\bf j}={({\bf B}\cdot \nabla \times {\bf B}-{\bf E}\cdot \nabla \times {\bf E}){\bf B}+(\nabla \cdot {\bf E}){\bf E}\times {\bf B} \over B^2}.
\end{equation}
Equations (\ref{maxwell}, \ref{ohm}) form a system of evolutionary nonlinear partial differential equations -- FFE. FFE is Lorentz covariant.

The FFE Ohm's law was designed (Gruzinov 1999) to maintain the force-freedom: 
\begin{equation}\label{force}
\rho {\bf E}+{\bf j}\times {\bf B}=0,~~~~\rho \equiv\nabla \cdot {\bf E}.
\end{equation}
One can show that if the force-free constraint (\ref{force}) is satisfied at the initial time (via ${\bf E}\cdot {\bf B}=0$ at the initial time), the FFE time evolution will maintain the force-free constraint (with ${\bf E}\cdot {\bf B}=0$). 

Consider the realm of applicability of FFE. If the stress-energy tensor of the system (plasma plus electromagnetic field) is dominated by the electromagnetic part, the force-free constraint is automatically satisfied. But it is unclear in which conditions would the plasma arrange the fields into a force-free configuration instead of accelerating the charges up to the energy equipartition with the field.

The force-free constraint takes a simple form in particular inertial frames. For arbitrary electric and magnetic fields, ${\bf E}$ and ${\bf B}$, at a given spacetime event, one can always find inertial frames with ${\bf E}$ parallel to ${\bf B}$ at this event. In such frames, the force-free condition means that (i) the current ${\bf j}$ is along ${\bf B}$, (ii) the electric field ${\bf E}$ vanishes. We will use this formulation of the force-freedom in our discussion of pulsar magnetospheres.

Consider a pulsar with the following Crab-like nominal parameters. Radius of the star $R_0=10$ km, angular velocity $\Omega=200~{\rm s}^{-1}$, spin-down power $L=5\times 10^{38}~{\rm erg}/{\rm s}$. The nominal magnetic field at the pole $B_0=5.8\times 10^{12}$ G (for axisymmetric pulsar, $L\approx \mu ^2\Omega ^4/c^3$, $\mu$ is the dipole). The light cylinder radius is $R_{\rm lc}\equiv c/\Omega=1500$ km. The nominal magnetic field at the light cylinder $B_{\rm lc}=B_0(R_{\rm lc}/R_0)^{-3}=1.7\times 10^6$ G. The nominal particle density near the star is $n_0\equiv\Omega B_0/(2\pi ce)=1.3\times 10^{13}~{\rm cm}^{-3}$ (Goldreich, Julian 1969, derived from the typical charge density required by stationary FFE, \S3). The nominal particle density near the light cylinder is $n_{\rm lc}=3.8\times 10^{6}~{\rm cm}^{-3}$.

The first force-free condition -- current flows along ${\bf B}$ -- must be satisfied. Even near the light cylinder, the cyclotron cooling time of electrons and positrons, $(3mc)/(4r_e^2B_{\rm lc}^2)=9\times 10^{-5}$ s ($r_e$ is the classical electron radius), is much smaller than the pulsar period, $2\pi/\Omega=3.1\times 10^{-2}$ s. Most of the charges must be in the lowest Landau state, with the residual perpendicular current corresponding to magnetization $\sim n_{\rm lc}\mu _B=3.5\times 10^{-14}$ G, $\mu _B$ is the Bohr magneton. We see that for any possible multiplicity (the ratio of true to nominal particle density), the perpendicular current is negligibly small. 

Actually the leading cause of current deviation from the magnetic field lines is the curvature of the magnetic field lines \footnote{I thank Jonathan Arons and Anatoly Spitkovsky for explaining this to me.}. To estimate this effect, one must calculate the ratio of the centrifugal pressure of the outflowing plasma to the magnetic pressure. Near the light cylinder, this reduces to the ratio of energy densities. The nominal value of this ratio, calculated assuming mildly relativistic outflow with small multiplicity, is $\sim  n_{\rm lc}mc^2/B^2_{\rm lc}=1.1\times 10^{-12}$. This must remain small for any reasonable multiplicity and Lorentz factor of the outflowing plasma, because the plasma outflow with high Lorentz factor would produce more curvature radiation than what is actually observed in the Crab pulsar (see below). 

The second force-free condition -- zero electric field along the magnetic field -- is more stringent. One can give a simple theoretical proof that the parallel electric field vanishes for radii $\lesssim 1000$ km. But this is not good enough to claim that FFE is quantitatively correct. We need FFE to work all the way to few light cylinder radii. Even though we do not know how exactly it happens, we will show, using the observational upper bounds on TeV Crab emission, that the parallel electric field must nearly vanish near the light cylinder too. 

First consider the $\lesssim 1000$ km zone. Here a clear-cut mechanism is available that would screen out the parallel electric field (Sturrock 1971, Ruderman, Sutherland 1975): (i) parallel electric field accelerates electrons and positrons, (ii) accelerated charges emit curvature photons, (iii) curvature photons produce electron-positron pairs on the background magnetic field, (i) parallel electric field accelerates electrons and positrons... -- leading to an avalanche that screens out the parallel electric field. We do not claim that this mechanism does operate in real pulsars. We only point out that this mechanism kills the parallel electric field, if the field is not actually killed by some other more powerful mechanism. 

For the numerical estimate, suppose that FFE does not apply starting from radii $\gtrsim R$. Then the parallel electric field at radius $R$ is close to the characteristic FFE value, $E\sim (\Omega R/c)B$, where the characteristic magnetic field at radius $R$ is $B\sim B_0(R/R_0)^{-3}$. This electric field will accelerate electrons and positrons to such Lorentz factors $\gamma$, that the curvature radiation power balances the electric power, $e^2c\gamma ^4/R^2\sim ceE$, giving the terminal Lorentz factor $\gamma \sim (2\pi n_0R_0^3)^{1/4}=10^8$. Electrons and positrons moving at such Lorentz factors emit curvature photons of energy $\epsilon \sim \gamma ^3c\hbar /R$. These photons will produce electron-positron pairs on the magnetic field $B$ with the mean free path $\lesssim R$, provided $\chi \equiv (\epsilon B)/(2mc^2B_q)>0.07$, where $B_q\equiv m^2c^3/(e\hbar )=4.4\times 10^{13}$G. This gives the maximal radius for the  pair production avalanche $R<780$ km. 

Now consider the zone around the light cylinder. Again assume that FFE does not apply. Then the parallel electric field will accelerate all available charges to Lorentz factors $\gamma =10^8$. These charges will emit curvature photons of energy $\epsilon _{\rm lc}\sim \gamma ^3c\hbar /R_{\rm lc}=0.13$ TeV. Since we know that FFE does apply just inside the light cylinder, the density of the charges near the light cylinder cannot be much smaller than $\sim  n_{\rm lc}$, and therefore the pulsed Crab emission at $\sim 0.1$ TeV cannot be much smaller than $\sim R_{\rm lc}^3n_{\rm lc}eB_{\rm lc}c\sim L=5\times 10^{38}~{\rm erg}/{\rm s}$. This is in gross contradiction with observations. The upper bound on Crab's pulsed emission at 100 Gev is about $\sim 3 \times 10^{33}~{\rm erg}/{\rm s}$, assuming 2 kpc distance (Oser et al 2001, extrapolation of Aharonian et al 2004). 

In summary: the current flows along the magnetic field, the parallel electric field vanishes. This means that the plasma is force-free, and the dynamics of the electromagnetic field reduces to FFE.

\section{Pulsar FFE and Pulsar Emission}

In covariant form, the force-free constraint (with ${\bf E}\cdot {\bf B}=0$) gives $F^{\mu \nu}j_\nu=0$, where $F$ is the electromagnetic field tensor, and $j$ is the 4-current. It follows that the electromagnetic stress-energy tensor is conserved, $\partial _\nu T^{\mu \nu}=0$, and  FFE conserves energy ${\cal E}$, momentum ${\bf P}$,  and angular momentum ${\bf M}$
\begin{equation}\label{cons}
{\cal E}\equiv \int d^3r {B^2+E^2\over 2}, ~~~ {\bf P}\equiv \int d^3r~{\bf E}\times{\bf B},~~~{\bf M}\equiv \int d^3r~{\bf r}\times ({\bf E}\times{\bf B}).
\end{equation}
The stationary (that is co-rotating) FFE magnetosphere of the pulsar of angular velocity ${\bf \Omega}$ is the stationary point of free energy (Gruzinov 2006) 
\begin{equation}\label{freeE}
{\cal E}'\equiv {\cal E} -{\bf \Omega}\cdot {\bf M}= \int d^3r (~{B^2+E^2\over 2}-{\bf \Omega}\cdot {\bf r}\times ({\bf E}\times{\bf B})~)
\end{equation}
under arbitrary variations $\delta {\bf E}$ of the electric field and iso-topological variations $\delta {\bf B}$ of the magnetic field 
\begin{equation}\label{isotop}
\delta {\bf B}=\nabla \times (\delta \vec{\xi }\times{\bf B}).
\end{equation}

Performing the variations, one gets the pulsar magnetosphere equation:
\begin{equation}\label{basic}
{\bf B}\times \nabla \times \left( {\bf B} + {\bf V}\times ({\bf V}\times{\bf B})\right) =0.
\end{equation}
Here ${\bf V}\equiv {\bf \Omega}\times {\bf r}$ is the co-rotation velocity, and the electric field is ${\bf E}=-{\bf V}\times {\bf B}$ \footnote{Which gives the Goldreich-Julian charge density used in \S2.}. The magnetosphere rotates at angular velocity ${\bf \Omega}$. The actual time evolution probably does take the magnetosphere into this ``minimum energy'' state (Spitkovsky 2006). 

The fact that the stationary magnetosphere equation (\ref{basic}) follows form the {\it iso-topological} variational principle has two important consequences. 

Stationary magnetospheres, that is solutions of (\ref{basic}), exist for any prescribed topology of the magnetic field lines. Pure FFE cannot choose among these topologies. In particular, one can have arbitrary current flowing in the closed-field zone. True, this current must dissipate, but it might be driven by the kinematic processes which produce the plasma. FFE cannot predict {\it the } pulsar power and magnetosphere, until the plasma production mechanism is understood quantitatively. 

Another consequence of the iso-topological minimization is the existence and stability of the singular current/charge layer along the magnetic separatrix (Gruzinov 2006). The singular current is known to exist in the axisymmetric magnetosphere (Contopoulos, Kazanas, Fendt 1999, Gruzinov 2005). The singular current layer is also seen in the generic inclined magnetosphere (Spitkovsky 2006).

For the axisymmetric pulsar, assuming pure dipole on the surface of the star and assuming zero current in the closed zone, one can calculate the surface current and charge on the magnetic separatrix numerically with about 10\% accuracy. For simplicity, consider only the region well within the light cylinder, $r\ll 1$, $z\ll 1$, where $r$ is the cylindrical radius and $z$ is the vertical cylindrical coordinate; we use natural ``pulsar units'' ($\Omega =\mu =c=1$).

The magnetic separatrix is a surface given by $r^2/z^3\approx 1.25$. The Goldreich-Julian charge density inside the separatrix  is $\rho=2B=4/z^3$. The current density on the magnetic axis (the line $r=0$) is $j\approx 3.8/z^3$, comparable to to $\rho$.  The current density decreases for growing $r$; the current reverses near the separatrix, but the total reverse current flowing inside the separatrix is small. There is no poloidal current outside the separatrix. Most of the reverse current flows in the singular current layer on the separatrix. The surface density of the separatrix current is $i\approx 1/r\approx (0.8r)/z^3$. 

The surface charge of the separatrix, $\sigma \approx 0.2r$, is much smaller than the surface current. This means that the separatrix current is formed by the counter-propagating surface flows of electrons and positrons. 

We will assume that the pulsar emission is primarily due to the anomalous Joule heating, and compare the Joule power of the separatrix and the Joule power of the bulk inside the separatrix. Since FFE was shown to be a good approximation, the width of the current layer on the separatrix $\delta$ due to kinetic non-FFE processes should  remain small, $\delta \ll r$. The current density in the separatrix layer $j_s\sim (r/\delta)j_b\gg j_b$, where $j_b$ is the current density in the bulk. The ratio of the Joule heating in the bulk to the Joule heating in the current layer 
\begin{equation}\label{joule}
{Q_b\over Q_s}\sim {E_bj_br^2\over E_sj_sr\delta}\sim {E_b\over E_s},
\end{equation}
where $E$ denote parallel electric fields in the bulk and in the separatrix current layer. For any ``reasonable'' anomalous resistivity, the electric field in the separatrix current layer should be much greater than the electric field in the bulk, because $j_s\gg j_b$. Then 
\begin{equation}\label{joule'}
{Q_b\over Q_s}\sim {E_b\over E_s}\ll 1.
\end{equation}
Given that the Joule heating in the bulk is much smaller than the Joule heating in the separatrix current layer, we propose that the pulsar emission is dominated by the separatrix current layer.

\acknowledgements
I thank Jonathan Arons and Anatoly Spitkovsky for discussions. This work was supported by the David and Lucile Packard foundation.

\end{document}